\documentclass[twocolumn,secnumarabic,amssymb, nobibnotes, aps, prl,superscriptaddress]{revtex4-1}

\def\vector#1{{\vec  #1}}

\def\kvec{\vector{k}}

\newcommand{\be}{\begin{equation}}
\newcommand{\ee}{\end{equation}}

 \usepackage{graphicx}
 \usepackage{epstopdf}
\usepackage{stackengine}

\begin{document}

\title{Line of Dirac Nodes in Hyper-Honeycomb Lattices}
\author{Kieran Mullen}
\affiliation{Department of Physics and Astronomy, University of Oklahoma, Norman, Oklahoma 73069, USA}
\author{Bruno Uchoa}
\affiliation{Department of Physics and Astronomy,  University of Oklahoma, Norman, Oklahoma 73069, USA}
\author{Daniel~T.~Glatzhofer}
\affiliation{Department of Chemistry and Biochemistry,  University of Oklahoma, Norman, Oklahoma 73069, USA }
\begin{abstract}

We propose a family of  structures that have ``Dirac loops", closed lines of Dirac nodes in momentum space, on which the density of states vanishes linearly with energy.  Those lattices all possess the planar trigonal connectivity present in graphene, but are three dimensional.  We show that  their highly anisotropic and multiply-connected Fermi surface leads to quantized Hall conductivities in three dimensions for magnetic fields with toroidal geometry. In the presence of spin-orbit coupling, we show that those structures have topological surface states.  We discuss the feasibility of realizing the structures as new allotropes of carbon.

\end{abstract}

\maketitle

{\it Introduction.}$-$ In honeycomb lattices, the existence of the Dirac point results from the planar trigonal connectivity of the sites and its sub-lattice symmetry \cite{GrapheneReview}.   Less well known are  ``Dirac loops",  three dimensional (3D) closed lines of Dirac nodes in momentum space, on which the energy vanishes linearly with the perpendicular components of momentum \cite{Balents}.   
To date there are no experimental observations of Dirac loops, and they were predicted to exist only in  topological superconductors \cite{Zhang} and 3D Dirac semimetals \cite{Wan} in which the parameters such as interactions and magnetic field are finely tuned \cite{Balents}. 

Theoretically, graphene is not the only possible lattice realization with planar trigonally connected atoms \cite{note0}.   It is therefore natural to ask if  there are variations on the honeycomb geometry that might produce exotic Fermi surfaces with Dirac-like excitations and topologically non-trivial states.  
 In this Letter, we propose a family of trigonally connected 3D lattices that admit simple tight-binding Hamiltonians having Dirac loops, without requiring any tuning or spin-orbit coupling.  
Some of these structures lie in the family of harmonic honeycomb lattices, which have been  studied in the context of the Kitaev model \cite{Kitaev, Kimchi, Mandal, Kim, Hermanns}, and experimentally realized in honeycomb iridates \cite{Iridates}. The simplest example is the hyper-honeycomb lattice, shown in Fig. 1a.

\begin{figure}[b]
 \centering
 \vspace{-0.5cm}
\includegraphics[width=2.7in]{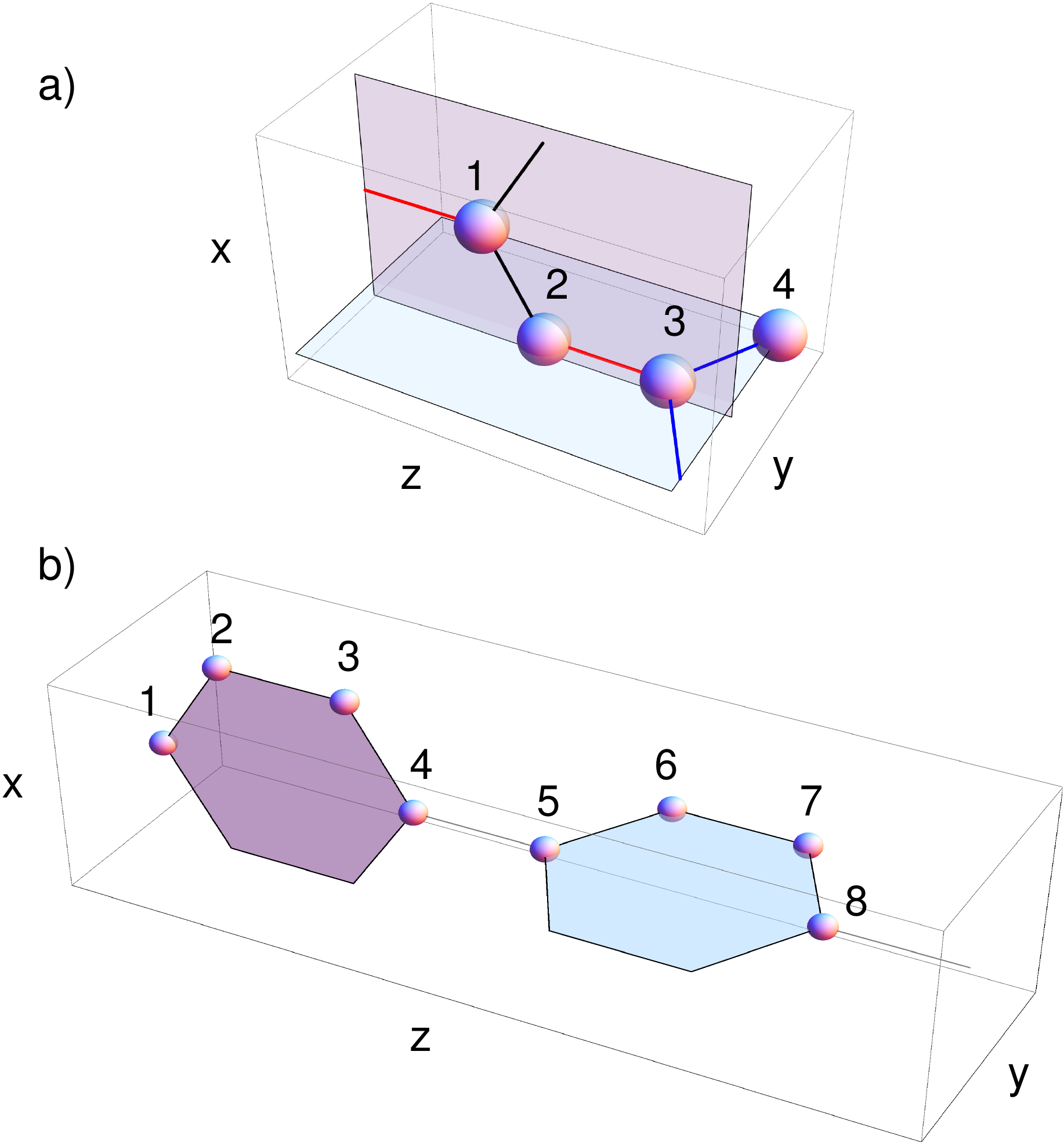}

  \caption{{\small (Color online) Simple lattice structures where all atoms are connected by three co-planar bonds spaced by $120^\circ$. a) The hyper-honeycomb lattice ($\mathcal{H}$-0), with a four atom unit cell. Atoms 1, 2 and 3 ($xy$ plane); atoms 2, 3 and 4 ($yz$ plane). Atoms 1 and 2 form a vertical chain  (black links);  atoms 3 and 4  form horizontal chain  (blue links).  The chains are connected by links (red) in the z-direction. b) An eight atom unit cell ($\mathcal{H}$-1).   Atoms 1-4 create a vertical chain of hexagons along the $x$-direction. Atoms 5-7 create a horizontal chain when repeated in the $y$-direction.  }} 
 \label{fig:polyacene}
 \end{figure}

We derive the low energy Hamiltonian of this family of systems, and analyze the quantization of the conductivity and possible surface states. Even though these systems are 3D semimetals,  their Fermi surface is multiply connected, with the shape of a torus, and highly anisotropic.  When a magnetic field with toroidal geometry is applied, we find that the Hall conductivity is quantized in 3D at sufficiently large field.  
Additional spin-orbit coupling effects can create topologically protected surface states in these crystals.  We claim that in the presence of spin-orbit coupling, these structures conceptually correspond to a new family of strong 3D topological insulators \cite{Hasan,Qi}.  We finally discuss the experimental feasibility of realizing those structures as new allotropic forms of carbon.

{\it Tight-binding lattice}.$-$ Our discussion starts with the simplest structure, the hyper-honeycomb lattice (see Fig. 1a). All atoms form three coplanar bonds spaced by $120^\circ$. The tight binding basis  
is of the form $\psi_{\alpha,\mathbf{k}}(\mathbf{r}) = \phi_\alpha(\mathbf{k}) \, e^{i \mathbf{k} \cdot \mathbf{r}}$,  with  $\alpha=1,2,3,4$ labeling  the components of a four vector $\Phi_\mathbf{k}$, which describes the amplitudes of the electronic wavefunction  on
the four atoms in the unit cell. The tight binding Hamiltonian satisfies the eigenvalue equation 
$
 { \mathcal{H}}  \Phi_\mathbf{k} = E  \Phi_{\mathbf{k}}
\label{eq:Htb}
$
where  
$
{\cal H}_{\alpha,\beta}=  t\sum_{\vec \delta_{\alpha,\beta}}   e^{i \mathbf{k}\cdot \vec\delta_{\alpha,\beta}}
$
and $t$ is the hopping energy between nearest neighbors sites separated by
the vector $\vec \delta_{\alpha ,\beta}$ connecting  an atom of the kind $\alpha$ with its nearest neighbor of the kind $\beta$.  The sum is carried over all nearest neighbor vectors $\vec\delta_{\alpha,\beta}$ among any two given species of sites, $\alpha$ and $\beta$. In explicit form,
\be
\mathcal{H}_{\alpha\beta}=t\left(\begin{array}{cccc}
0 & \Theta_{x} & 0 & \mbox{e}^{-ik_{z}a}\\
\Theta_{x}^{*} & 0 & \mbox{e}^{ik_{z}a} & 0\\
0 & \mbox{e}^{-ik_{z}a} & 0 & \Theta_{y}\\
\mbox{e}^{ik_{z}a} & 0 & \Theta_{y}^{*} & 0\end{array}\right)
 \label{hmat}
\ee
where $\Theta_{i}=2\mbox{e}^{ik_{z}a/2}\cos(\sqrt{3}k_{i}a/2)$ with
$i=x,y$  and $a$ the interatomic distance.

This Hamiltonian has a zero energy eigenvalue  along the curve defined by $k_z=0$ and
\be
4 \cos{ \left( { \sqrt{3}} k_x a/2\right) }  \cos{\left( { \sqrt{3}} k_y a/2\right) }=1.
\label{nodalLine}
 \ee
 Eq. (\ref{nodalLine}) defines a zero energy line $\mathbf{k}_0=(k_x(\phi),k_y(\phi),0)$ shown in the solid white lines in Fig. 2a, where $\phi$ is the cylindrical polar angle with respect to the center of the Brillouin zone (BZ) at the $\Gamma$ point. The reciprocal lattice is generated by the vectors $\mathbf{b}_1=(2\pi/\sqrt{3}a,0,\pi/3a)$, $\mathbf{b}_2=(0,2\pi/\sqrt{3}a,-\pi/3a)$ and $\mathbf{b}_3=(0,0,2\pi/3a)$, as shown in Fig. 2b, and has four high symmetry points, $\Gamma,\,R,\,X,$ and $Z$. The 3D BZ has four-fold rotational symmetry around the $[001]$ direction. 
 The energy spectrum of  Hamiltonian (1) has four bands, shown in Fig. 2c, where the two lowest energy bands are particle hole-symmetric and cross along the nodal lines, in the $k_z=0$ plane. The bands displayed in Fig 2c follow the path shown in the triangular line of panels a, b, with the point $R$ located in the middle of the flattened corners of the BZ. 
 
 \begin{figure}[t]
 \centering
\includegraphics[width=3.2in]{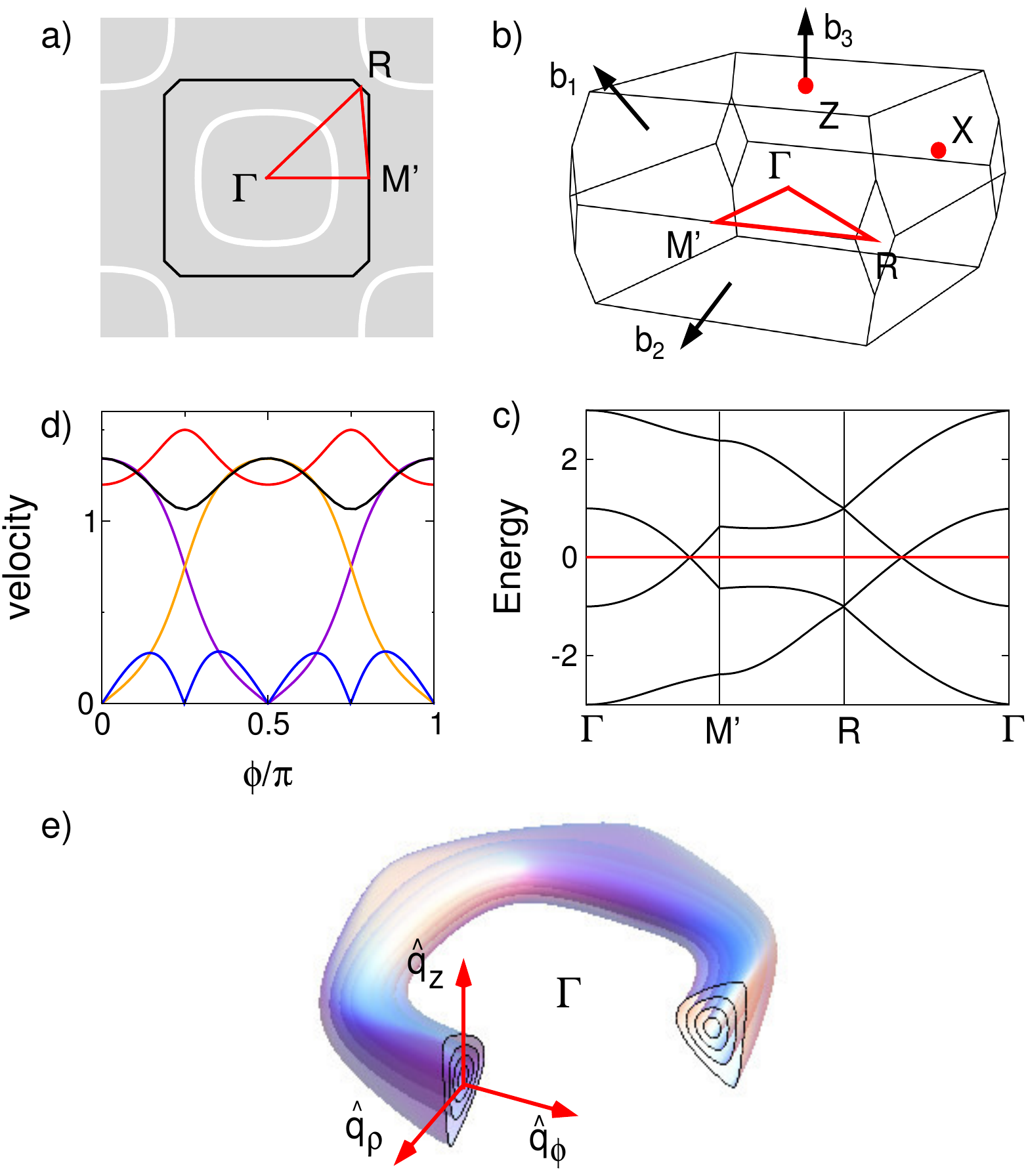}
  \caption{{\protect\small (Color online) a) BZ in the  $k_z=0$ plane showing the Dirac loop lines (solid white). Black line: boundary of the BZ, centered at the $\Gamma$ point.  b) 3D Brillouin zone. Black arrows: directions of the reciprocal lattice vectors $\mathbf{b}_i$, $i=1,2,3$. c) Energy spectra of the four bands of Eq. (\ref{hmat}) in units of $t$ plotted  along the path shown in the red line of panels a) and b). The low energy bands cross along the Dirac loop.   d) Velocity of the quasiparticles at the Dirac line in units of $ta$, as a function of the cylindrical polar angle $\phi$ with respect to $\Gamma$. In cylindrical coordinates, red: $v_z(\phi)$; black: $v_\rho(\phi)$; blue: $v_\phi(\phi)$. The orange and violet curves describe $v_x(\phi)$ and  $v_y(\phi)$ respectively. e) Red arrows: cylindrical moving basis around the line of Dirac nodes. Toroidal Fermi surfaces for energies  $E/t= 0.1,\,0.2,\, 0.3$ and $0.4$ around the Dirac loop.   }} 
 \label{fig:spectra}
 \end{figure} 
 
 {\it Projected Hamiltonian}.$-$  Expanding the $\Phi$ eigenvectors around the nodal line and projecting the Hamiltonian (\ref{hmat}) in the two component  subspace that accounts for the lowest energy bands, the projected Hamiltonian can be written in the Dirac-like form
 \be
\mathcal{H}_{\rm p}(\mathbf{q}) = - \left[  v_x(\phi)  q_x + v_y(\phi) \,  q_y\right] \sigma_x + v_z(\phi) q_z \, \sigma_z\,,
\label{proj}
\ee
where $\mathbf{q}\equiv \mathbf{k}(\phi)-\mathbf{k}_0(\phi)$ is the momentum measured away from the nodal line,  $\sigma_x$, $\sigma_z$ are $2\times2$ Pauli matrices (we set $\hbar\to 1$) and 
\begin{equation}
\pm E_\mathbf{k}=\pm\sqrt{[v_x(\phi)  q_x + v_y(\phi) \,  q_y]^2 + [v_z(\phi) q_z]^2  }
\end{equation}
is the low energy spectrum. The quasiparticles of Hamiltonian (\ref{proj}) are  chiral in that there is a Berry phase  $ i \oint \langle  \Phi_\mathbf{k} \vert \vec \nabla_\mathbf{k}  \Phi_\mathbf{k}\rangle \cdot d\kvec = \pi$ \cite{Xiao, Volovik} associated with paths in momentum space that encircle the nodal line.

The Fermi  velocities $v_i(\phi)$ ($i=x,y,z$) are plotted in Fig. 2d, and can be approximated by simple trigonometric functions. The quasiparticles disperse linearly in the normal directions to the nodal line (Fig 2e) and are dispersionless along the Dirac loop. In the cylindrical moving basis shown in Fig. 2e, the velocities are given by $v_z(\phi)$, $v_\rho(\phi)$ and $v_\phi(\phi)$. Even though the nodal line is not a perfect circle,  the ratio $v_\phi(\phi)/v_z(\phi)$ is small and oscillates between 0 and $0.19$.   Away from half-filling, the Fermi surfaces are toroids containing the nodal line $\mathbf{k}_0(\phi)$, as shown in Fig. 2e.  For small energies, the cross-section is nearly circular, and the energy varies linearly with the distance from the loop.
A similar analysis can be done for the unit cell shown in Fig. 1b, which has 8 carbon atoms in the unit cell. In that case, the tight binding Hamiltonian is an 8$\times$8 matrix with 8 different bands. This Hamiltonian can be projected into the low energy states, resulting in a Hamiltonian with the same form as Eq. (\ref{proj}). 

 The above structures are merely two in a hierarchy of possible lattices that can be made with perpendicular  zigzag chains of trigonally connected carbon atoms.  
We denote these structures with two integers  $(n_x, n_y)$, where $n_x$ ($n_y$) is the number of vertical (horizontal)  complete honeycomb hexagons  contained in the unit cell.  In this notation, the hyper-honeycomb lattice shown in Fig.\ref{fig:polyacene}a describes a 
$(0,0)$ lattice, while Fig. \ref{fig:polyacene}b has one complete honeycomb hexagon along both the vertical and horizontal zigzag chains in the unit cell, and hence is a $(1,1)$ structure. The symmetric higher order structures $(n,n)$ belong to the family of harmonic honeycomb lattices, denoted as $\mathcal{H}$-$n$, with $n\in \mathbb{N}$ \cite{Iridates}. In this family,  the screw axis symmetry is preserved and they all display Dirac loops at zero energy around the $\Gamma$ point, with the $\mathcal{H}$-0 case shown in Fig. 1a being the simplest atomic chain arrangement. In the $n\to\infty$ limit, those structures describe a single layer of graphene. 
Asymmetric structures where $n_x\neq n_y$  have a very anisotropic unit cell and their nodal lines are displaced in the BZ. 
   
The simplest Hamiltonian that captures the physics described in Hamiltonian (\ref{proj}) is a  minimal model where we approximate the nodal line (\ref{nodalLine}) by a circle with the average radius $k_0\equiv \langle k_\rho(\phi) \rangle \approx 1.61 a^{-1}$. The in-plane velocity is independent of the angle $\phi$,
\be
\mathcal{H}_0(\mathbf{q})  = -v_\rho  q_\rho \sigma_x + v_z  q_z \sigma_z\,,  \label{eq:Htoy}
\ee
where $q_\rho = k_\rho - k_0$, is a small variation in the cylindrical radial momentum $k_\rho^2\equiv(k_x^2 + k_y^2)$ away from the radius of the nodal line,   and $v_\rho\equiv \langle v_\rho(\phi)\rangle\sim 1.22 ta$, and $v_z\equiv \langle v_z(\phi)\rangle\sim 1.32ta$,  are the average  velocities in the $\hat{\rho}$ and $\hat{z}$  directions. When the nodal line is a perfect circle, $v_\phi=0$. 
 The density of states per volume varies linearly with the energy 
$
D(E)=  k_0 E/(2\pi v_\rho v_z)  \label{eq:DoS}
$, 
including a factor of two for the spin degeneracy. 
 
{\it Charge transport.}$-$ 
For short range impurities, the DC conductivity can be calculated self-consistently  at zero temperature \cite{note}.    Going back to the projected Hamiltonian (\ref{proj}), we define the Green's function $\hat{G}_\mathbf{k}(\omega) = [\omega - \mathcal{H}_\mathrm{p}(\mathbf{k})- \hat \Sigma(\omega)+i0^+]^{-1}$, where the 2$\times$2 matrix $\hat \Sigma(\omega) = V_0^2 \sum_\mathbf{k} \hat G_\mathbf{k}(\omega)$ is the self-energy due to a local quenched disorder potential $V_0$ 
\cite{QuantCondCalc}.  

At zero frequency, the self consistent solution of the self-energy $\hat \Sigma(0)=i\Gamma$ is diagonal and purely imaginary, with $\Gamma$  the scattering rate. In the minimal model (\ref{eq:Htoy}), $\Gamma \approx t \,\mbox{exp}\{- 1/[V_0 D(V_0)]\}$.  The DC conductivity in the direction $\hat n$ is
$
\sigma_{\hat n}(0) =  e^2 
\mathrm{tr} \sum_{\mathbf{k}}  \hat v_{\hat n} \hat A(\mathbf{k},0)  \hat v_{\hat n}\hat A(\mathbf{k}, 0) 
$
, where  $\hat A(\mathbf{k},0) = -2 \mathrm{Im} \hat{G}_\mathbf{k}(0)= -2\Gamma/(E_\mathbf{k}^2+\Gamma^2)$ is the static spectral function, $e$ is the electron charge and $\hat v_{\hat n}=\hat{n}\cdot \nabla_\mathbf{q} \mathcal{H}_\mathrm{p}$
is the velocity operator projected along the $\hat n$ direction.  

In 3D, the conductivity has units of $e^2/h$ divided by length (restoring $\hbar$) \cite{Vish, Aji}.
When $\Gamma \ll t$, the conductivity is independent of the scattering rate,  as expected \cite{QuantMinCond1},  and gives 
\begin{equation}
\sigma_{\hat n}(0)=   \frac{C_{\hat n}}{a}  \frac{e^{2}}{\pi h}  ,\label{eq:sigmaz-1}
\end{equation}
per spin, where  $C_{\hat n}$ is a {\it non-universal} dimensionless geometrical factor. In the $\mathcal{H}$-0 lattice,  \begin{equation} 
C_z\approx1.79,\qquad C_x=C_y\approx0.75. \label{eq:C}
\end{equation}  
In the minimal model (\ref{eq:Htoy}), $C_z=k_0 a v_z/v_\rho\sim 1.76$, while $C_\rho=3k_0 a v_\rho/(8v_z)\sim 0.55$  for transport along any direction in the plane of the Dirac loop, in qualitative agreement with (\ref{eq:C}). Those values contrast with the theoretical conductivity (per spin) of Dirac fermions in 2D for unitary disorder, $\sigma(0) = e^2/(\pi h)$ \cite{QuantMinCond1, QuantMinCond2}.

{\it 3D Quantum Hall effect.}$-$ In the presence of a uniform magnetic field,  the minimal model  (\ref{eq:Htoy})  becomes 
$
\mathcal{H}_0(\rho,z)  = -v_\rho   \sigma_x(i \partial_\rho- A_\rho)+ v_z  \sigma_z (i\partial_z - A_z), 
$ where $\mathbf{A}= A_z\hat z +A_\rho \hat \rho+ A_\phi \hat \phi$
 is the vector potential. In this model, a toroidal magnetic field $B_\phi \hat \phi$  pointing along the Dirac loop corresponds in the symmetric gauge to a vector potential 
$\mathbf{A}=-B_{\phi}\rho\hat{z}$.  Such a field can be created with a  time dependent electric field applied along the $\hat z$ direction. 
Taking the square of the Hamiltonian (\ref{eq:Htoy}), 
$
\mathcal{H}_0^2(\xi) = (\sqrt{v_\rho v_z}/\ell_B)  [ (\xi^2 -\partial_\xi^2 )\sigma_0 + \sigma_y ],
$
where $\sigma_0$ is the identity matrix,
$\xi\equiv\sqrt{v_{z}/v_{0}} \left(\rho/\ell_{B}-k_{z}\ell_{B}\right)$ is a dimensionless coordinate 
and $\ell_{B}=\sqrt{h/B_{\phi}e}$
is the magnetic length.  Rewriting the Hamiltonian in terms of ladder operators of the 1D Harmonic oscillator, $a=(\xi+\partial_\xi)/\sqrt{2}$ and $a^\dagger=(\xi-\partial_\xi)/\sqrt{2}$,  the energy spectrum has a zeroth LL and is the same as in conventional 2D Dirac fermions in a magnetic field,
\begin{equation}
E_N= \mathrm{sign}(N) (\sqrt{2v_\rho v_z}/\ell_B)  \sqrt{|N|},\label{eq:E4}
\end{equation}
where $N\in\mathbb{Z}$. 

Although being a 3D semimetal, in a perfectly circular Dirac loop, the system has a 3D quantum Hall effect \cite{Halperin, Koshino, Bernevig} at {\it any} magnetic field $B_\phi$. In more conventional field geometries, the Hall conductivity of 3D crystals was shown by Halperin \cite{Halperin} to be in the form $\sigma_{ij} = e^2/(2\pi h) \epsilon_{ijk} G_k$, where $\epsilon_{ijk}$ is the antisymmetric tensor and $\mathbf{G}$ is a multiple of some reciprocal lattice vector (it could also be zero). Following the TKNN analysis \cite{TKNN}, a necessary and sufficient requirement for quantized Hall conductivities in general is that the band structure will open insulating {\it bulk} gaps at finite applied magnetic field, and that the Fermi level will lie in one of those gaps. In 3D, the quantum Hall effect has been observed or predicted before only in systems with extreme anisotropies \cite{Balicas, McKernan}, or else in strongly anisotropic systems with Dirac quasiparticles, such as Bernal stacked graphite \cite{Bernevig}, which are more easily susceptible to LL quantization. 
 
In the toroidal geometry of the magnetic field, each Dirac cone in the loop contributes with $(N+\frac{1}{2})e^2/h$ quanta per spin to the Hall conductivity.  In cylindrical coordinates, for a circular nodal line,
\begin{eqnarray} \sigma_{z \rho} &=&  2\frac{e^2}{h} \int \frac{\mathrm{d}k^3}{(2\pi)^3} \mathrm{Im} \left \langle \partial_{k_\rho} \Phi_{\mathbf{k}} | \partial_{k_z} \Phi_\mathbf{k} \right \rangle \nonumber\\
&=& (2N+1)\frac{e^2}{h} \int \frac{\mathrm{d} \phi}{2\pi}  k_0 = (2N+1)k_0 \frac{e^2}{h},
 \end{eqnarray}
accounting for the spin degeneracy 2, while $\sigma_{\rho \phi } =0$.  Hence, in the presence of a $B_\phi \hat \phi$ field, a radial current along the $\hat \rho$ direction creates a voltage difference along the $\hat z$ direction and vice versa. By adiabatic continuity, the Hall conductivity is invariant under deformations of the nodal line (up to trivial scaling effects), provided that the LL gaps do not close completely.  Hence, nodal lines with nearly circular shape will show quantized Hall conductivities in the toroidal field geometry whenever the Fermi level lies in the energy gap, at  finite magnetic field.  
This property opens the prospect in the future for the observation of 3D quantum Hall effect in other classes of systems that prove to have Dirac loops as well. 
 
{\it Topological surface states.}$-$  In the presence of spin-orbit coupling effects, the surface states can acquire topological character. 
The spin orbit coupling can be included through a trivial generalization of the Kane-Mele model \cite{Kane,Fu} for the $\mathcal{H}$-$n$ lattice, 
$
\mathcal{H}_{ij}^{\mathrm{SO}}=\sum_{l}it_{2}(\mathbf{d}_{i l}\times\mathbf{d}_{ l j})\cdot\vec{\tau} 
$, where $ij$ are next-nearest neighbor sites connected by two nearest neighbor vectors $\mathbf{d}_{il}$,  $\vec{\tau}=(\tau_{x},\tau_{y},\tau_{z})$ is a vector
of Pauli matrices acting in the spin space, and $t_2 = \Delta_{\mathrm{SO}}/(3\sqrt 3)$ gives the spin-orbit coupling gap. 
The Rashba coupling is detrimental to the spin-obit coupling gap, but is expected to be small when mirror symmetry in the plane of the atomic bonds is preserved.  In the $\mathcal{H}$-0
lattice, the total Hamiltonian is an $8\times8$ matrix in the $\Phi_\mathbf{k}$ basis. In the general case,  the Hamiltonian of the $\mathcal{H}$-$n$ structure is a matrix with $2^{3+n}\!\times 2^{3+n}$ components, including the spin. 

In Fig. 3, we show the energy spectrum of the $\mathcal{H}$-0 and $\mathcal{H}$-1 crystals  in the presence of a spin-orbit coupling $t_2=0.1t$. We considered the geometry of an infinite slab oriented along the $y$ direction, with surfaces along the $x$ and $z$ ones. The modes that cross zero energy are surface states at the two [100] surfaces of the crystals.  All structures have two helical spin polarized modes per surface, which cross at the center of the BZ, at the $\Gamma$ point. Those surface modes are topologically protected by Kramers theorem, and describe a new possible family of strong 3D topological insulators. Due to the four-fold symmetry of the BZ,  identical surface states can also be found in the two [010] surfaces for a slab geometry rotated around the $\hat z$ axis by $\pi/2$. The [001] surfaces, nevertheless, do not have those states. 

\begin{figure}[t]
 \centering
\includegraphics[width=3.2in]{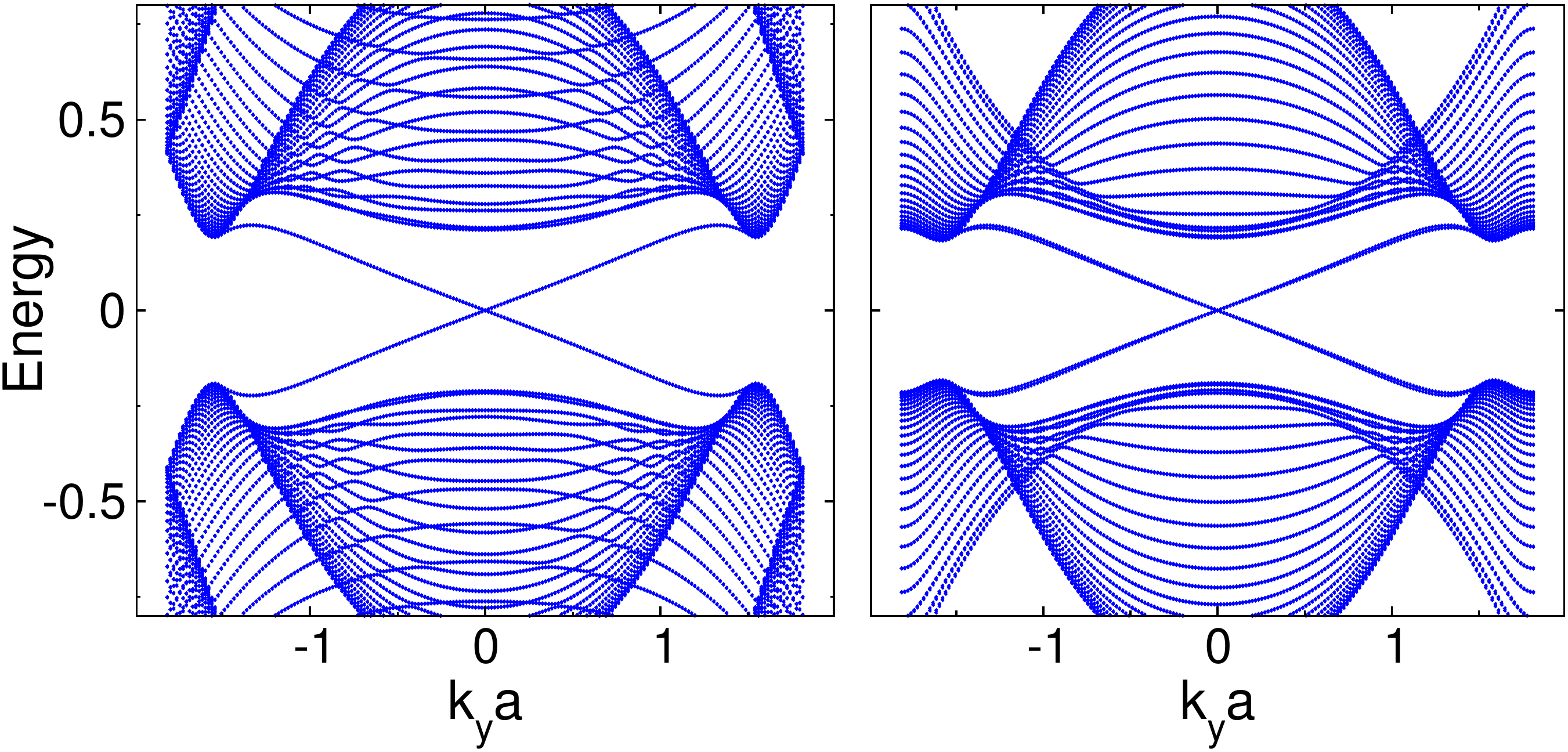}

  \caption{{\protect\small (Color online) Energy bands in units of $t$ in the presence of a large spin-orbit coupling  $t_2=0.1t$ (see  text).   The crossed lines at zero energy are topological {\it surface} states along the [100] direction of the crystal. Left: $\mathcal{H}$-0 crystal; right: $\mathcal{H}$-1 crystal.    }} 
 \label{fig:fermi}
 \end{figure}

{\it Synthesis as a new carbon allotrope.}$-$ Due to $\pi-\pi$ orbital  interactions between the chains, $\mathcal{H}$-$n$ lattices could   likely be realized as metastable allotropic forms of carbon \cite{Hoffman}. The planar trigonal bonding of the carbon atoms is nevertheless quite robust.  Simulations with Tersoff potentials \cite{note4} indicate that hyper-honeycomb allotropes of carbon atoms could be as stable, or even more stable than other metastable allotropes such as diamond. 

Although synthesis of this new family of carbon allotropes can  be challenging, the  $\mathcal{H}$-0 allotrope could be synthetized in a layer by layer fashion using mono-functionalized carbon chains of atoms in the alkyne or alkynide groups \cite{glatz10}. Those groups can be coordinated  perpendicularly to a surface, in a way as to allow epitaxial polymerization in the form of a monolayer of oriented chains \cite{glatz20}. Once the first layer is grown, the exposed functional groups can be replaced with a new layer of functionalized chains perpendicular to the first one \cite{glatz30}. The subsequent repetition of those two stages can lead to a 3D lattice of carbon atoms deposited as a film on the substrate surface. 
A similar method can be applied for instance to the $\mathcal{H}$-1  allotrope \cite{glatz40}, as possibly to the entire family of harmonic structures. 

The realization of topological surface states in those carbon allotropes can be very difficult due to the smallness of the spin-orbit gap, which is of the order of 0.1 meV ($t_2 \sim 10^{-4}t$), as in graphene \cite{Min}. Nevertheless, a substantial enhancement of the gap can be achieved by chemically doping those structures with adatoms such as thallium (Tl) \cite{Weeks}.  In graphene, Tl adatoms are expected the create a spin-orbit gap of the order of 20meV ($t_2\approx 0.02t$) while keeping the planar trigonal bonds of carbon intact and the Rashba coupling parametrically small.  We speculate that a similar enhancement of the spin-orbit gap is possible in the 3D structures as well, and will be considered somewhere else.

{\it Acknowledgements.}$-$ B.U. thanks A. Jaefari and Y. Barlas for discussions. K. M. was supported by NSF grant DMR-1310407. B. U. acknowledges NSF CAREER grant DMR-1352604 for support.

{\it Note.}$-$ During the preparation of this version of the manuscript, we became aware of the recent experimental observation of a line of Dirac nodes in Ca$_3$P$_2$ \cite{Xie} and of a related works on inversion symmetric crystals \cite{Kim2} and graphene networks \cite{Weng}, which appeared after our original preprint.



\begin{thebibliography}{99}
\bibitem{note} Time reversed paths that encircle the Dirac line pick up a relative phase of $\pi$ and therefore cancel. However, backscattering between opposite sides of the BZ across the nodal line  is not suppressed, as they can be described by paths that pick an overall Berry phase of $2\pi$.
\end{thebibliography}
\end{document}